\documentclass[journal,onecolumn]{IEEEtran}
\ifCLASSINFOpdf
\else
\fi

\usepackage{graphics} 
\usepackage{subcaption}
\usepackage[pdftex]{graphicx}
\usepackage{amssymb} 
\usepackage{xspace}
\newtheorem{theorem}{Theorem}
\newtheorem{corollary}{Corollary}
\newtheorem{lemma}{Lemma}

\newtheorem{definition}{Definition}
\newcommand{\losslessDelay}{{lossless-delay}\xspace}
\newcommand{\worstCaseDelay}{{worst-case-delay}\xspace}
\newcommand{\oracleInfo}{side information\xspace}
\usepackage{color, colortbl}
\newcommand{\delay}{\tau}
\newcommand{\len}{t}
\newcommand{\burst}{b}

\newcommand{\BuildingBlock}{Building block\xspace}

\newcommand{\spreadvgms}{Spread Code\xspace}
\newcommand{\learningCode}{Spread ML Code\xspace}

\newcommand{\replaceStar}{{j_*}}
\newcommand{\replaceI}{{i_*}}

\newcommand{\parityOffline}{p^{(IP)}}

\newcommand{\algModified}{Algorithm 2\xspace}

\newcommand{\x}{X}
\newcommand{\y}{Y}

\newcommand{\oracle}{O}
\newcommand{\s}{S}
\newcommand{\p}{P}
\newcommand{\pV}{P^{(v)}}
\newcommand{\pVSize}{\parity}

\newcommand{\uPacket}{U}
\newcommand{\vPacket}{V}
\newcommand{\maxSize}{m}
\newcommand{\sample}{\mathcal{S}}

\newcommand{\regret}{\mathcal{R}}

\newcommand{\rate}{R^{(\E)}}
\newcommand{\rateOptimal}{R^{(\E,Opt)}}
\newcommand{\regretVariable}{z}

\newcommand{\policy}{\mathcal{F}}
\newcommand{\policyOffline}{f^{(IP)}}

\newcommand{\policyShort}{f}
\newcommand{\policyPredictShort}{f^{(\epsilon)}}
\newcommand{\E}{\mathrm{E}}
\newcommand{\var}{\mathrm{Var}}

\newcommand{\state}{\mathcal{X}}

\newcommand{\parity}{p}
\newcommand{\parityOther}{p^{\dagger}}

\usepackage{dsfont}
\newcommand{\Equation}{Eq.}
\newcommand{\Equations}{Eqs.}
\newcommand{\identity}{\mathds{1}}
\usepackage{float}
\usepackage{dsfont}
\floatstyle{ruled}
\newfloat{alg}{thp}{scratch}
\floatname{alg}{Algorithm}
\newcommand{\onlineOptimal}{online-optimal-rate\xspace}
\newcommand{\offlineOptimal}{offline-optimal-rate\xspace}
\newcommand{\messagePacket}{message packet\xspace}
\newcommand{\messagePackets}{message packets\xspace}
\newcommand{\channelPacket}{channel packet\xspace}
\newcommand{\channelPackets}{channel packets\xspace}
\newcommand{\messagePacketSizeSequence}{message packet size sequence\xspace}

\usepackage{pgfplots}
\usepackage[utf8]{inputenc}

\newcommand{\F}{\ensuremath{\mathbb F}}

\usepackage[T1]{fontenc}

\usepackage{ifthen}
\usepackage{cite}
\usepackage{float}
\usepackage{dsfont}
\floatstyle{ruled}
\newfloat{alg}{thp}{scratch}
\floatname{alg}{Algorithm}
\usepackage{afterpage}
\usepackage{url}

\usepackage[cmex10]{amsmath}

\pgfplotsset{compat=1.15}
\begin{document}

\title{Learning-Augmented Streaming Codes are Approximately Optimal for Variable-Size Messages}

\author{Michael Rudow and K.V. Rashmi
\thanks{M. Rudow and K.V. Rashmi are with the Computer Science Department, Carnegie Mellon University, Pittsburgh,
PA, 15213 USA.}
\thanks{This is an extended version of the IEEE ISIT 2022 paper~\cite{rudow2022learning} with the same title}
}
%
%

%

\maketitle


\begin{abstract}

Real-time streaming communication requires a high quality-of-service despite contending with packet loss. 
Streaming codes are a class of codes best suited for this setting. 
A key challenge for streaming codes is that they operate in an ``online'' setting in which the amount of data to be transmitted varies over time and is not known in advance. 
Mitigating the adverse effects of variability requires spreading the data that arrives at a time slot over multiple future packets, and the optimal strategy for spreading depends on the arrival pattern. Algebraic coding techniques alone are therefore insufficient for designing rate-optimal codes.
We combine algebraic coding techniques with a learning-augmented algorithm for spreading to design the first approximately rate-optimal streaming codes for a range of parameter regimes that are important for practical applications.


\end{abstract}

\section{Introduction}
\label{sec:introduction}

Real-time communication arises in many popular applications, including VoIP, online gaming, and videoconferencing. 
These applications involve a sender transmitting packets of information to a receiver over a lossy channel. 
The receiver must decode the data within a strict playback deadline. 
In many scenarios, one cannot retransmit the lost packets because doing so requires an extra round trip time and can thus exceed the maximum tolerable latency~\cite{badr2017perfecting}. 
Instead, one can use erasure coding to recover lost packets. 

\begin{figure}[t]
    \centering
   \includegraphics[width=.65\columnwidth]{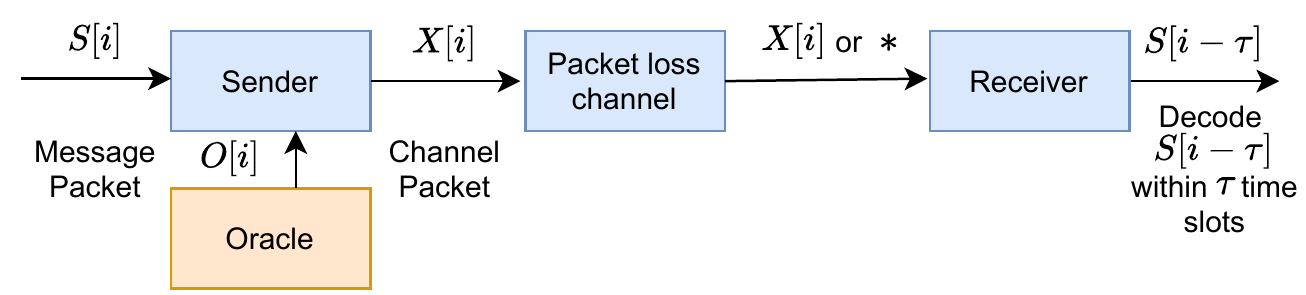}
    \caption{Overview of the model for streaming codes.}
    \label{fig:model}
\end{figure}

While erasure coding has been well studied, real-time communication has several unique aspects that require a new ``streaming model,'' as was introduced by Martinian and Sundberg~\cite{martinian2004burst}. 
During each time slot, $i$, a ``\messagePacket,'' denoted as $\s[i]$, of size $k$ arrives at a sender. 
The sender then transmits a ``\channelPacket,'' denoted as $\x[i]$, of size $n$ to a receiver over a burst-only loss channel. 
The sender must recover $\s[i]$ by time slot $(i+\delay)$. 
An overview of the model is presented in Figure~\ref{fig:model}, with the sender, channel, and receiver appearing in blue. (The component ``\oracleInfo'' is introduced later.) 
Coding schemes that recover lost symbols from each \messagePacket $\delay$ time slots later have significantly higher rates than {alternatives, such as} interleaved maximal distance separable codes, that recover all lost symbols together by $\delay$ time slots after the \textit{first} \messagePacket for which the symbols are lost~\cite{martinian2004burst}. 

Numerous works~\cite{martinian2007delay,badr2017layered,fong2018optimalLong,krishnan2018rate,krishnan2020rate,domanovitz2019explicit,badr2018multiplexed,fong2020optimalM,badr2011diversity,badr2015streamingM,haghifam2021streaming,adler2017burst,leong2012erasure,leong2013coding,badr2015streaming,Fong2020optimal,Badr2017FEC,li2011correcting,wei2013prioritized,krishnan2019simple,ramkumar2020staggered,su2021random} have employed tools from algebraic coding theory to design optimal streaming codes for various settings where the sizes of \messagePackets and \channelPackets are fixed in advance. 
These regimes are suitable for applications that involve sending fixed quantities of data, such as VoIP when audio packets are sent uncompressed. 

In contrast, many applications, such as videoconferencing, involve transmitting \textit{variable} amounts of data. 
A new streaming model incorporating \messagePackets of varying sizes was introduced in~\cite{rudow2018variable}. 
Two factors affect the optimal rate in this setting. First is the sequence of the sizes of all \messagePackets of a transmission, called 
the ``\messagePacketSizeSequence.'' Second is the maximum number of time slots the receiver can wait to decode each \messagePacket under a lossless transmission, called ``lossless delay'' and denoted as $\delay_L$. 

The optimal rate for ``offline'' schemes (which know the \messagePacketSizeSequence) exceeds that of ``online'' schemes (which cannot access the sizes of \messagePackets for future time slots) in all but two settings~\cite{rudow2020online}. 
In one setting where $\delay_L$ has its maximum possible value, a technique for spreading the symbols of each \messagePacket over several \channelPackets independently of all other \messagePackets is rate optimal. The other setting, where $\delay_L$ has its minimum possible value (i.e., $0$), requires sending the symbols of each \messagePacket in the corresponding \channelPacket. Therefore, information about the sizes of the future \messagePackets does not help. 
Rate-optimal constructions, or even approximately rate-optimal constructions, are not known even for the offline setting for all remaining parameter regimes. 

{We consider the setting of $\delay_L=1$, which is important since it is the smallest value for which the symbols of a \messagePacket can be spread over \textit{multiple} \channelPackets. 
Spreading helps to significantly mitigate the adverse effect of variability of the sizes of \messagePackets on the rate~\cite{rudow2018variable}. 
On the other hand, maintaining a small value of $\delay_L$ is crucial for latency-sensitive applications, as the delay of $\delay_L$ extra time slots may be incurred for decoding \textit{every} \messagePacket. 
}

{We first consider the offline setting and decompose the code design into two distinct challenges. 
First, how can we best spread message symbols over \channelPackets? 
Second, how can we send the minimum necessary number of parity symbols to ensure that each \messagePacket is decoded in time, given any choice of how to spread message symbols? 
We use an integer program offline to determine how to optimally spread message symbols over \channelPackets. 
We then introduce a building block for constructing a rate-optimal streaming code for \textit{any given} choice of how to spread message symbols over \channelPackets. 
One final challenge remains: how can we construct a rate-optimal \textit{online} streaming code? 
}

We address the problem by \textit{combining machine learning with tools from algebraic coding theory.} 
We take a learning-based approach, relying on techniques similar to empirical risk minimization to convert the optimal offline solution into an approximately optimal online one that maximizes the expected rate. Our proposed method determines how to spread symbols online, and then the building block construction is applied.

Our methodology can be viewed as using a ``learning-augmented algorithm''\textemdash a topic that has recently surged to prominence, tackling problems in other domains, such as caching~\cite{lykouris2018competitive}, metric task systems~\cite{antoniadis2020online}, bloom filters~\cite{Mitzenmacher2018model}, learned index structures~\cite{kraska2018case}, scheduling~\cite{Lattanzi2020online}, etc.~\cite{bamas2020primal,mitzenmacher2020algorithms,hsu2019learning,Jiang2020Learning-Augmented,anand2020customizing}. However, to the best of our knowledge, the powerful paradigm of learning-augmented algorithms has not been applied to design coding schemes until now.

Using machine learning models to perform encoding and/or decoding has received considerable attention in the recent past, including for channel coding~\cite{gruber2017on,nachmani2018deep,lugosch2017neural,dorner2018deep,cammerer2017scaling,jiang2019turbo,jiang2019deepturbo,nachmani2019hyper,makkuva2021ko}, decoding with feedback~\cite{kim2018deepcode,kurka2020deep}, approximate coded computation~\cite{kosaian2019parity,kosaian2020learning,krishna2020collage}, MIMO~\cite{jeon2017blind,samuel2019learning}, etc.~\cite{farsad2017detection,gui2018deep,oshea2017an,oshea2016learning,choi2018necst}. Most of these works use neural networks to handle encoding and or decoding in a black-box manner. 
In contrast, our work applies machine learning only to a small portion of the problem: to {determine} how to spread message symbols to address the uncertainty in the sizes of the future \messagePackets. 
{We then leverage} tools from classical algebraic coding theory to solve the rest of the problem. This makes the learning part lightweight and interpretable and allows {for} theoretical guarantees on the rate.

\afterpage{
\begin{figure*}[t]
    \centering
\includegraphics[width=\linewidth]{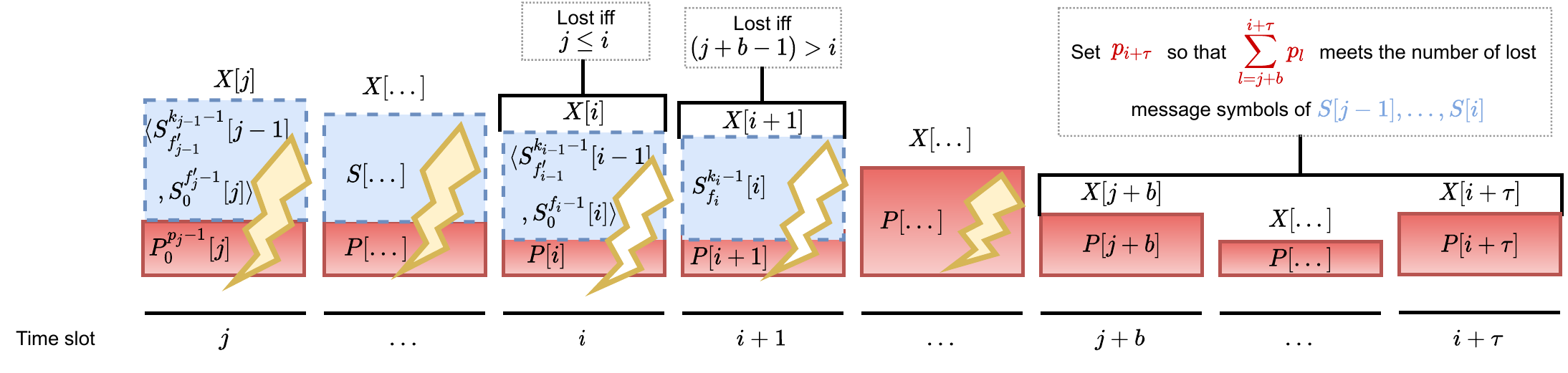}
    \caption{Selecting $p_{i+\tau}$ by considering each burst starting in time slot $j \in \{i-\burst+1,\ldots,i+1\}$ (shown with lightning bolts). } 
    \label{fig:codeParity}
\end{figure*}}
\section{Model and background}
\label{sec:backgroundAndmodel}

We now present the {system} model used in this work, which is {built on top of} the model of streaming codes for variable-size messages~\cite{rudow2020online,rudow2018variable}.

A transmission occurs over $(\len+1)$ time slots for a non-negative integer $\len$. 
During the $i$th time slot for $i \in \{0,\ldots,\len\}$, the sender obtains a \messagePacket, $\s[i] \in \F^{k_i}$, where $\F$ is a finite field, and $k_i \in \{0,\ldots,\maxSize\}$ for a positive integer $\maxSize$. 
The sender also receives ``\oracleInfo,'' $\oracle[i]$, that  captures the differences between the online and offline settings. 
In the offline setting, which assumes knowledge of the future, the \oracleInfo is the sizes of the future \messagePackets. 
In the online setting, the side information is independent $\sample$ samples from the distribution of the sizes of future \messagePackets for some positive integer $\sample$. 
Let $D_{k_0,\ldots,k_i}$ be the conditional distribution of $k_{i+1},\ldots,k_\len$ given $k_0,\ldots,k_i$. Then,
\begin{equation}
\label{eq:defOracle}
\oracle[i] = \begin{cases}
    \left(k_{i+1},\ldots,k_{\len}\right) & \text{if offline}\\
    \big \langle \left(k_{i+1}^{(j)},\ldots,k_{\len}^{(j)}\right) \sim D_{k_0,\ldots,k_i} \mid j \in \{0,\ldots,\sample-1\} \big \rangle    
    & \text{if online}. 
\end{cases}
\end{equation}

During the $i$th time slot, the sender transmits a \channelPacket, $\x[i] \in \F^{n_i}$ to a receiver, where $n_i$ is a non-negative integer. 
The receiver obtains $\y[i] \in \{\x[i], *\}$, which reflects packet reception or packet loss, respectively. 
When all \channelPackets are received, $\s[i]$ must be decoded by the receiver by time slot $(i+\delay_L)$ for a parameter $\delay_L \ge 0$. 
This requirement is called the ``\losslessDelay'' constraint and represents the maximum tolerable latency for lossless transmission. 
{Recall that our work considers $\delay_L = 1$ (as discussed in Section~\ref{sec:introduction}).} 
In addition, $\s[i]$ must be decoded by time slot $(i+\delay)$ when losses occur for a parameter $\delay \ge \delay_L$. 
This reflects the maximum acceptable latency in the worst case and is called the ``\worstCaseDelay'' constraint.

Our work uses the packet loss model from previous work~\cite{martinian2004burst, martinian2007delay, rudow2020online}. 
Packets are lost in bursts of up to $\burst$ consecutive losses followed by at least $\delay$ successful receptions. 
Formally, for any $i \in \{0,\ldots,\len -\delay-\burst+1\}$, if $\left(\y[i] = *\right) $ then $\forall j \in \{i+\burst,\ldots,i+\burst+\delay-1\}$
$\left( \y[j] = \x[j] \right).$ 
To ensure that the \worstCaseDelay constraint is satisfiable, $\delay \ge \burst$. 
Furthermore, $\burst>0$ because coding is not needed otherwise. 
Finally, $\delay > \burst$, since $\delay \ge (\delay_L+\burst)$~\cite{rudow2018variable} and $\delay_L = 1$ in our work.

When the sizes of \messagePackets and \channelPackets are fixed, the rate is simply the ratio of their sizes. However, a more nuanced notion of rate is needed due to the variability in the sizes of \messagePackets and \channelPackets. The rate for a \messagePacketSizeSequence $k_0,\ldots,k_\len$ is defined as 
\begin{equation}
    R_\len = \frac{\sum_{i=0}^\len k_i}{\sum_{i=0}^\len n_i}.
\end{equation}

For any integer $i$, $\{0,\ldots,i\}$ is denoted by $[i]$. 
For convenience of notation, $k_i = 0$ for $i \in [2\delay-1] \cup \{\len-2\delay+1,\ldots,\len\},$ and $\len$ is taken to be at least $4\delay$. This assumption is satisfied by adding zero padding, which does not affect the rate. As such, $\s[0],\ldots,\s[2\delay-1]$ are known to be {of size} zero, and  $\x[0],\ldots,\x[2\delay-1]$ are empty. 
Encoding and decoding depends on the history of the transmission, which is summarized as follows.
\begin{definition}[State]
For any $\len,\delay,$ and $i \in \{2 \delay,\ldots,\len\}$, the \textit{state} is denoted $\state_i = (k_0,\ldots,k_i, \x[0],\ldots,\x[i-1])$.
\end{definition}
This section considers systematic codes for clarity, but the results also hold for general codes. 
{To meet the \losslessDelay constraint, the symbols of $\s[i]$ must be sent by time slot $(i+\delay_L)$ (i.e., in $\x[i]$ and $\x[i+1]$).  
The ``policy'' of a construction, as defined below, specifies how to spread the message symbols. }
\begin{definition}[Policy]
\label{def:policy}
The \textit{policy} of a construction for any $i \in [\len]$ and state $\state_i$ is the number of symbols of $\s[i]$ sent in \channelPacket $\x[i]$. 
The policy is denoted as $\policy_i\left(\state_i\right)$ (or $\policyShort_i$ for conciseness) and lies in $[k_i]$. 
\end{definition}
For any $i > 0$, $\x[i]$ comprises (a) the first $\policyShort_i$ symbols of $\s[i]$, (b) the final $(k_{i-1} - \policyShort_{i-1})$ symbols of $\s[i-1]$, and (c) $\parity_i$ parity symbols, denoted as $\p[i]$. 
The encoding is given by $\x[i] =$
\begin{equation}
\label{eq:encode}
Enc(\state_i, \s[i-\delay],\ldots,\s[i-1],\s_0[i],\ldots,\s_{\policyShort_i-1}[i], \oracle[i])  
\end{equation}
for $i \ge 2\delay$, and $\x[i]$ is empty for $i < 2\delay$. 
{This section assumes that $\x[i]$ is independent of the message symbols of $\s[i]$ sent in $\x[i+1]$
for clarity, although the results hold without this assumption.} 
The receiver obtains $\y[i] \in \{\x[i], *\}$ 
depending on whether \channelPacket $\x[i]$ is received or dropped.\footnote{{The receiver needs the sizes of the \messagePackets to decode. Thus,} a small header with up to $2\sum_{j=i-\burst}^i\log(k_j) \le 2(\burst+1) \log(\maxSize)$ symbols containing $\left(k_{i-\burst},\ldots,k_i\right)$ is {added to the header of} $\x[i]$.} 
Under lossless transmission, $\s[i]$ is available in uncoded form. 
Otherwise, $\s[i]$ is decoded as
\begin{equation}
\label{eq:decode}
Dec\left(\left \langle \y[j], k_j, \policyShort_j \mid j \in [i+\delay]\right\rangle \right).    
\end{equation}

The following notation is used throughout our work. 
A vector $V$ has length $v$, comprises symbols $(V_0,\ldots,V_{v-1})$, and is a row vector. 
For any $i\le j \in [v-1]$, $V_i^j = (V_i,\ldots,V_j)$.

\section{A \BuildingBlock construction}
\label{sec:construction}
In this section, we present a rate-optimal construction, called the ``$\left(\delay, \burst,\len,\left\langle \policyShort_i \mid i \in [\len]\right\rangle\right)-$\spreadvgms,'' for any given policies, i.e., choice of how to spread the message symbols over \channelPackets. Specifically, for any given $\left\langle \policyShort_i \mid i \in [\len]\right\rangle$, at least $\policyShort_i$ symbols of $\s[i]$ will be sent in $\x[i]$ under the construction for each $i\in [\len]$.

The first $2\delay$ \channelPackets are empty. For each $i \in [\len] \setminus [2\delay-1]$, $\x[i]$ comprises (a) the first $\policyShort_i'$ symbols of $\s[i]$ for some $\policyShort_i ' \ge \policyShort_i$, (b) the final $(k_{i-1}-\policyShort_{i-1}')$ symbols of $\s[i-1]$, and (c) $\parity_i$ parity symbols called $\p[i]$. 
Next, we define $\policyShort_i '$, $\parity_i$, and $\p[i]$ for any \messagePacketSizeSequence, $k_0,\ldots,k_\len$.

\noindent \textbf{{Defining each $\policyShort_i'$ and $\parity_i$}.} 
\noindent {For time slots $i \in [2 \delay - 1] \cup \{\len-2\delay +1,\ldots,\len\}$, $\policyShort_i' = k_i = 0$.} 
For time slots $i \in [3\delay-1] \cup \{\len-\delay+1,\ldots,\len\}$, $ \parity_{i} = 0$. 
For all $i = 2\delay,\ldots,(\len-2\delay)$, we define $\parity_{i+\delay}$ {to be as small as possible while ensuring that $\s[i]$ is decoded by time slot $(i+\delay)$ under any lossy transmission. }  
Specifically, $\parity_{i+\delay} = $
\begin{equation}
\label{eq:paritySize}
    \max_{j \in \{i-\burst+1,\ldots,i+1\}} \Big(0,\identity[j+\burst-1 \ge i+1] \left(k_i- \policyShort_i \right)+ \identity[j \le i]\policyShort_i + \sum_{l = j}^{i} (k_{l-1}- \policyShort_{l-1}') + \sum_{l = j}^{i-1}\policyShort_{l}' - \sum_{l=j+\burst}^{i+\delay-1} \parity_l \Big),
\end{equation}
as is illustrated in Figure~\ref{fig:codeParity}. 
We then use $\parity_{i+\delay}$ to define
\begin{align}
\label{eq:adaptSpreadDef}
\policyShort_i ' = \max \left( \policyShort_i, \parity_{i+\delay}\right).
\end{align}

\begin{figure*}
    \centering
\includegraphics[width=.85\linewidth]{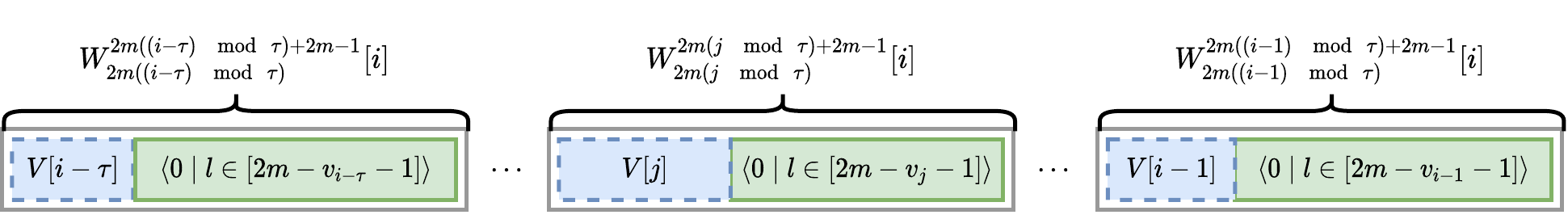}
    \caption{Defining $W[i]$ by placing the symbols of $\vPacket[i-\tau]$ in positions $2m (j \bmod \delay),\ldots,\left(2m (j \bmod \delay)+v_j-1\right)$ for $j \in \{i-\tau,\ldots,i-1\}$. The remaining positions are filled with $0$'s.}
    \label{fig:W}
\end{figure*}

\noindent \textbf{Constructing parity symbols.} The parity symbols are defined analogously to {those of the construction from \cite{rudow2020online} (which builds on the construction from~\cite{badr2017layered}).} 
For $i \in \{2\delay, \ldots, \len-\delay\}$, {the message symbols sent in} $\x[i]$ are partitioned into (a) symbols of $\s[i]$ that are recovered during time slot $(i+\delay)$ under a lossy transmission, and (b) symbols of $\s[i-1]$ and $\s[i]$ that are recovered by time slot $(i-1+\delay)$ under a lossy transmission. 
The two components are of sizes $u_i$ and $v_i$ and are denoted as $\uPacket[i]$ and $\vPacket[i]$, respectively. 
Thus, $\x[i] = \left(\uPacket[i],\vPacket[i],\p[i]\right)$, where
\begin{align}
 \uPacket[i]&=\s_0^{\parity_{i+\delay}-1}[i] \label{eq:defU}\\
\vPacket[i] &= \big(\s_{\parity_{i+\delay}}^{k_i-\policyShort_{i}'-1}[i], \s_{\policyShort_{i-1}'}^{k_{i-1}-1}[i-1]\big)\label{eq:defV}\\
\p[i] &=  \uPacket[i-\delay]+ \pV[i]\label{eq:defP}.
\end{align}
Each symbol of $\pV[i]$ is a linear combination of the symbols of $\left(\vPacket[i-\delay],\ldots,\vPacket[i-1]\right)$, where the linear equations are chosen using a $(2\maxSize \delay) \times (2\maxSize \delay)$ Cauchy matrix, $A$, as follows. Let $W[i]$ be a length $2 \maxSize \delay$ vector where positions $2m(j \bmod \delay),\ldots,\left(2m(j\bmod\delay) +2m-1\right)$ comprise $\vPacket[j]$ followed by $(2m-v_j)$ $0$'s for $j \in \{i-\delay,\ldots,i-1\}$, as is illustrated in Figure~\ref{fig:W}. 
Finally, we define 
\begin{equation}
    \label{eq:defW}
    \pV[i] = W[i]A_{(i)},
\end{equation}
where $A_{(i)}$ is $A$ restricted to columns $2m(i \bmod \delay),\ldots,\left(2m(i \bmod \delay) + \parity_i-1\right)$. 
The $\left(\delay, \burst,\len,\left\langle \policyShort_i \mid i \in [\len]\right\rangle\right)-$\spreadvgms is shown recovering a burst in Figure~\ref{fig:spreading}. 

\noindent \textbf{Decoding.} For $i \in [\len]$, $\s[i]$ is decoded (a) from $\x[i]$ and $\x[i+1]$ under lossless conditions, and (b) by solving a system of linear equations corresponding to the symbols of $\s[i-\delay],\ldots,\s[i-1],\y[i],\ldots,\y[i+\delay]$ when losses occur.

\afterpage{
\begin{figure*}
    \centering
\includegraphics[width=.85\linewidth]{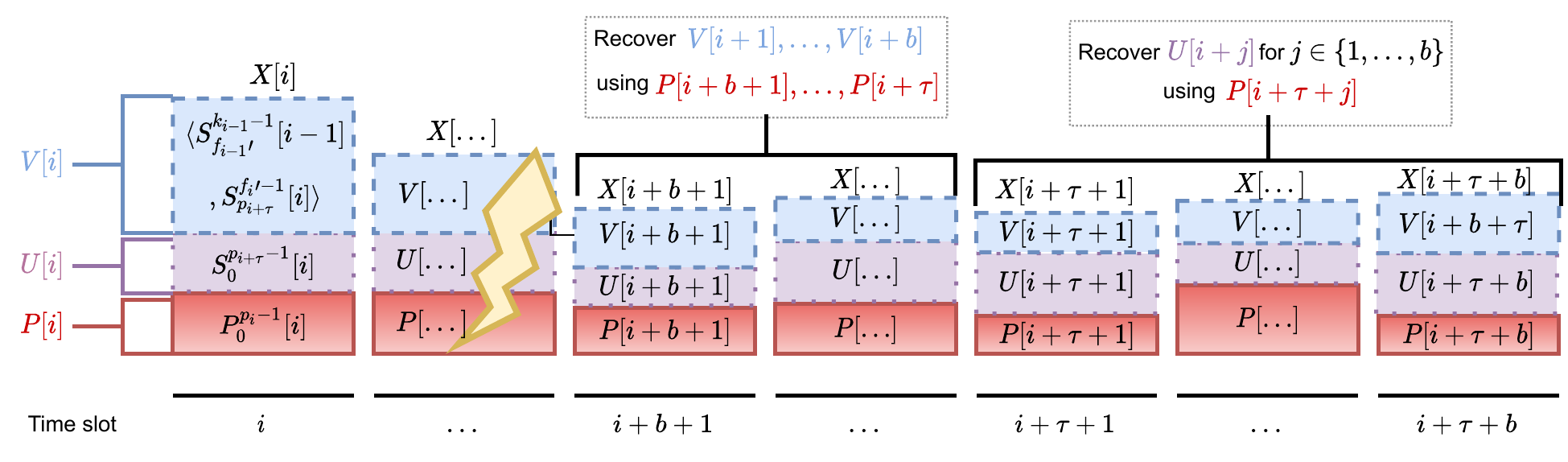}
    \caption{An example of how the $\left(\delay, \burst,\len,\left\langle \policyShort_i \mid i \in [\len]\right\rangle\right)-$\spreadvgms recovers a burst of length $\burst$ starting in time slot $(i+1)$.} 
    \label{fig:spreading}
\end{figure*}
}
Next, we show that the $\left(\delay, \burst,\len,\left\langle \policyShort_i \mid i \in [\len]\right\rangle\right)-$\spreadvgms {\xspace meets the \losslessDelay and \worstCaseDelay constraints}. 

\begin{lemma}
\label{lem:correctSpreadVGMS}
For any parameters $(\delay, \burst$), an arbitrary \messagePacketSizeSequence $k_0,\ldots, k_\len$, and any policy $\policyShort_i$ for $i \in [\len]$, the  $\left(\delay, \burst,\len,\left\langle \policyShort_i \mid i \in [\len]\right\rangle\right)-$\spreadvgms satisfies the \losslessDelay constraint and \worstCaseDelay constraint.
\end{lemma}

\begin{IEEEproof}
For any $i \in [\len-1]$, $\s[i] = \left(\s_0^{ \policyShort_i' - 1}[i], \s_{ \policyShort_{i}'}^{k_{i}-1}[i]\right)$, where $\s_0^{ \policyShort_i' - 1}[i]$ is sent in $\x[i]$, and $\s_{ \policyShort_{i}'}^{k_{i}-1}[i]$ is sent in $\x[i+1]$. 
For $\s[\len]$, $k_{\len}=0$ is known.
Thus, the \losslessDelay constraint is met. 

Next, we show that the \worstCaseDelay is satisfied for any burst. 
Satisfaction is immediate if the burst starts after $(\len-\delay-\burst+1)$, since $k_{\len-\delay-\burst}=0,\ldots,k_\len=0$ is known. 
Otherwise, suppose $\x[i],\ldots,\x[i+\burst-1]$ are lost for some $i \in [\len-\delay-\burst+1]$. 
We assume that $i\ge 2\delay$, since $k_0=0,\ldots,k_{2\delay-1} = 0$ is known, and no symbols are sent in $\x[0],\ldots,\x[2\delay-1]$. 
Each $\pV[i+\burst] =  \left(\p[i+\burst]-\uPacket[i+\burst-\delay]\right),\ldots, \pV[i+\delay-1] = \left(\p[i+\delay-1] - \uPacket[i-1]\right)$ is known. 

We show that enough parity symbols are received after the burst by time slot $(i+\delay-1)$ to recover $\vPacket[i],\ldots,\vPacket[i+\burst-1]$  as follows: 
\begin{align}
\policyShort_{i+\burst-1} + \sum_{j=i}^{i+\burst-1} \left(k_{j-1} - \policyShort_{j-1}'\right)  + \sum_{j=i}^{i+\burst-2} \policyShort_j'  \le&  \sum_{j=i+\burst}^{i+\burst+\delay-1} \pVSize_j \label{eq:useparitySize0}  \\
\sum_{j=i}^{i+\burst-1} \left(k_{j-1} - \policyShort_{j-1}' + \policyShort_j'\right)  \le&  \sum_{j=i+\burst}^{i+\burst+\delay-1} \pVSize_j \label{eq:useparitySize}  \\
\sum_{j=i}^{i+\burst-1} v_j+ u_j \le&  \sum_{j=i+\burst}^{i+\burst+\delay-1} \pVSize_j\label{eq:defUV}\\
\sum_{j=i}^{i+\burst-1} v_j \le&  \sum_{j=i+\burst}^{i+\delay-1} \pVSize_j, \label{eq:sizeU}
\end{align}
where \Equation~\eqref{eq:useparitySize0} follows from \Equation~\eqref{eq:paritySize}, 
\Equation~\eqref{eq:useparitySize} follows from (a) $\policyShort_{i+\burst-1}' = \policyShort_{i+\burst-1}$, or (b) combining $\policyShort_{i+\burst-1}' = \parity_{i+\burst-1+\delay}$ with \Equation~\eqref{eq:paritySize} to show
$$ \sum_{j=i}^{i+\burst-1} \left(k_{j-1} - \policyShort_{j-1}'\right)  + \sum_{j=i}^{i+\burst-2} \policyShort_j' = k_{i-1} - \policyShort_{i-1}' + \sum_{j=i}^{i+\burst-2} k_j \le \sum_{j=i+\burst}^{i+\burst+\delay-2} \pVSize_j,$$
\Equation~\eqref{eq:defUV} follows from \Equations~\eqref{eq:defU} and~\eqref{eq:defV},  and \Equation~\eqref{eq:sizeU} follows from \Equations~\eqref{eq:defU} and ~\eqref{eq:defP}. 

Next, we show that $\p[i+\burst],\ldots,\p[i+\delay-1]$ suffice to recover $\vPacket[i],\ldots,\vPacket[i+\burst-1]$ 
Recall that $\pV[j] = W[j]A_{(j)}$ for $j \in \{i+\burst,\ldots,i+\delay-1\}$, where $W[j]$ contains $\vPacket[l]$ in positions 
$$I^{(i,j)} = \bigcup_{l\in \{i,\ldots,i+\burst-1\}}\{2m(l \bmod \delay),\ldots,\left(2m(l \bmod \delay) + v_l-1\right)\},$$
as defined in \Equation~\eqref{eq:defW} and illustrated in Figure~\ref{fig:W}. 
Let $W'[j]$ be the vector of length $2m\tau$ with (a) $0$'s in positions in $I^{(i,j)}$, (b) $W_r[j]$ for positions $r \in [2m\tau-1] \setminus I$. 
The receiver can compute $W'[j]$ and use it to determine $P^{*}[j] = \left(\pV[j] - W'[j]A_{(j)}\right)$. 
Let $l_0=i,\ldots,l_{\burst-1} = (i+\burst-1)$, and $r = \arg \min_{l \in \{i,\ldots,i+\burst-1\}} (l \bmod \delay)$. 
Let $l_0'=(i+\burst),\ldots,l_{\delay-\burst-1} = (i+\delay-1)$, and $r' = \arg \min_{l \in \{i+\burst,\ldots,i+\delay-1\}} (l \bmod \delay)$. 
Then 
\begin{equation}
    \begin{bmatrix}
    P^*[l_{r'-(i+\burst)}']^T \\
    \vdots \\
    P^*[l_{\delay-\burst-1}']^T \\
    P^*[l_{0}']^T \\
    \vdots \\
    P^*[l_{r'-(i+\burst)-1}']^T
    \end{bmatrix}^T = 
    \begin{bmatrix}
    \vPacket[l_{(r-i)}]^T \\
    \vdots \\
    \vPacket[l_{\burst-1}]^T \\
    \vPacket[l_{0}]^T \\
    \vdots \\
    \vPacket[l_{(r-i)-1}]^T
    \end{bmatrix}^T
    A_{(i)}',
\end{equation}
where $T$ means transpose, and $A_{(i)}'$ is a submatrix of a Cauchy matrix of dimensions $\left(\sum_{j=i}^{i+\burst-1} v_j\right) \times \left(\sum_{j=i+\burst}^{i+\delay-1} \pVSize_j\right)$. 
As such, $A_{(i)}'$ is full rank, allowing the receiver to solve for $\vPacket[i],\ldots,\vPacket[i+\burst-1]$.

Finally, for $j = i, \ldots,(i+\burst-1)$, the receiver uses the symbols of $\vPacket[j],\ldots,\vPacket[j+\delay-1]$ to compute $\pV[j+\delay]$, yielding $\uPacket[j] = \left(\p[j+\delay] - \pV[j+\delay]\right)$. 
As $\s[j]$ is sent over $\vPacket[j],\uPacket[j],$ and $\vPacket[j+1]$, it is recovered by time slot $(j+\delay)$.
\end{IEEEproof}

Next, we provide a lower bound on the number of parity symbols sent by any streaming code that satisfies the \losslessDelay and \worstCaseDelay constraints. 
The bound is illustrated in Figure~\ref{fig:systematicConstraints}
\afterpage{
\begin{figure*}[t]
    \centering
\includegraphics[width=\linewidth]{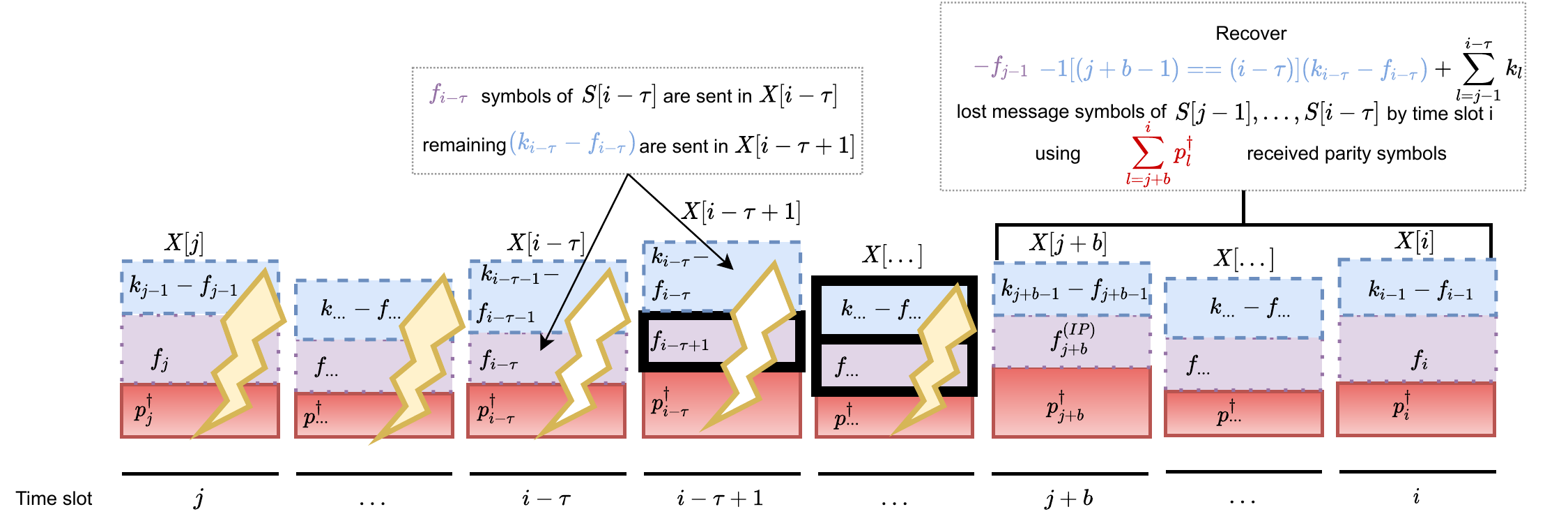}
    \caption{Illustration of the bound on the number of symbols sent under any streaming code satisfying the delay constraints. For any $i \in \{3\delay, \ldots, \len\}$ and $j \in \{i-\delay-\burst+1,\ldots,i-\delay+1\}$, $\s[j-1],\ldots,\s[i-\delay]$ are recovered by time slot $i$ when a burst of length $\burst$ starting in time slot $j$ (shown with lightning bolts),  under the relaxation of receiving the lost symbols of $\s[i-\delay+1],\ldots,\s[j+\burst-1]$ (boxes with thick black outline).  } 
    \label{fig:systematicConstraints}
\end{figure*}
}

\begin{lemma}
\label{lem:worstCaseConstraint}
{Consider any $\delay,\len,\burst$, and any streaming code that satisfies the \losslessDelay and \worstCaseDelay constraints. 
{Suppose for $l \in [\len]$, the construction} sends $\parityOther_{l}$ parity symbols and uses policy $\policyShort_l$.} 
For any $i \in \{3\delay, \ldots, \len\}$ and $j \in \{i-\delay-\burst+1,\ldots,i-\delay+1\}$, the number of parity symbols satisfies 
\begin{equation}
\label{eq:parityBound}
-\policyShort_{j-1}- \identity[j +\burst-1=i-\delay] \big(k_{i-\delay} - \policyShort_{i-\delay}\big)  + \sum_{l=j-1}^{i-\delay} k_l \le  \sum_{l=j+\burst}^{i} \parityOther_{l}.
\end{equation}
\end{lemma}

\begin{IEEEproof}
Suppose $\x[j],\ldots,\x[j+\burst-1]$ are lost. 
Due to the \worstCaseDelay constraint, $\s[j-1],\ldots,\s[i-\delay]$ must be recovered by time slot $i$. 
Thus, $\sum_{l=j-1}^{i-\delay} k_l$ symbols need to be decoded. 
Because the \messagePackets are independent, the symbols of $\x[0],\ldots,\x[j-2]$ contain no information about $\s[j-1],\ldots,\s[i-\delay]$. 
By definition of encoding (that is, \Equation~\eqref{eq:encode}), $\x[j-1]$ contains $\policyShort_{j-1}$ symbols of $\s[j-1]$, no additional information about $\s[j-1]$, and no information about $\s[j],\ldots,\s[i-\delay]$. 
When $(j+\burst-1) == (i-\delay)$, $\x[i-\delay+1]$ is received, and its message symbols include $(k_{i-\delay}-\policyShort_{i-\delay})$ symbols of $\s[i-\delay]$. 
The remaining message symbols of $\x[j+\burst],\ldots,\x[i]$ correspond to $\s[i-\delay+1],\ldots,\s[i]$ and cannot be used to recover $\s[j-1],\ldots,\x[i-\delay]$ (independence of \messagePackets). 
Altogether, 
\begin{equation*}
    -\policyShort_{j-1}- \identity[j+\burst-1 =i-\delay] \big(k_{i-\delay} - \policyShort_{i-\delay}\big) + \sum_{l=j-1}^{i-\delay} k_l
\end{equation*}
symbols corresponding to $\s[j-1],\ldots,\s[i-\delay]$ need to be recovered by time slot $i$. 
These symbols can only be recovered using the parity symbols of $\x[j+\burst],\ldots,\x[i]$, of which there are 
$$\sum_{l=j+\burst}^i \parityOther_l.$$ 

The symbols of \messagePackets are drawn uniformly at random from the underlying field. 
Thus, the total number of parity symbols must match the number of message symbols to be decoded. 
\end{IEEEproof}

{We show} the rate of the $\left(\delay, \burst,\len,\left \langle \policyShort_i \mid i \in [\len]\right\rangle\right)-$\spreadvgms matches that of any streaming code with policy $\policyShort_i$ for $i \in [\len]$.
\begin{lemma}
For any $\delay,\len,\burst$, and \messagePacketSizeSequence $k_0,\ldots,k_\len$, the $\left(\delay, \burst,\len,\left \langle \policyShort_i \mid i \in [\len]\right\rangle\right)-$\spreadvgms matches the rate of any streaming code with policy $\policyShort_i$ for $i \in [\len]$ that satisfies the \losslessDelay and \worstCaseDelay constraints.
\label{lem:changeSameRate}
\end{lemma}
\begin{IEEEproof}Under the  $\left(\delay, \burst,\len,\left \langle \policyShort_i \mid i \in [\len]\right\rangle\right)-$\spreadvgms, $\sum_{l=0}^{\len} (k_l+\parity_l)$ symbols are sent. 
Consider any streaming code construction that satisfies the \losslessDelay and \worstCaseDelay constraints, and for each $i \in [\len]$, employs policy $\policyShort_i$ and sends $\parityOther_i$ parity symbols. 
This streaming code construction sends $\sum_{l=0}^{\len} \left( k_l + \parityOther_{l}\right)$ symbols in total.

We show by induction on $i = 0, \ldots, \len$ that $\sum_{l=0}^{i} \parity_l \le \sum_{l=0}^i \parityOther_{l}$. 
The base case holds for $j < 3\delay$ because $\parity_{0} = 0, \ldots, \parity_{3\delay-1}=0$. 
For the inductive hypothesis, we note for all $j <i$ :
\begin{equation}
\label{eq:IHConvert}
\sum_{l=0}^{j} \parity_l \le \sum_{l=0}^{j} \parityOther_{l}.   
\end{equation}

In the inductive step, consider $i = 3\delay,\ldots,\len-\delay$. 
By \Equation~\eqref{eq:IHConvert}, the proof holds if $\parity_i \le \parityOther_{i}$. Otherwise, $\parity_i > \parityOther_{i} \ge 0$. 
Due to \Equation~\eqref{eq:IHConvert}, we only need to show for $j \le i$ that $\sum_{l=j}^i \parity_l \le \sum_{l=j}^i \parityOther_{l}$.

Let $\replaceI = (i-\delay)$. By \Equation~\eqref{eq:paritySize}, there exists $\replaceStar \in \{\replaceI-\burst+1,\ldots,\replaceI+1\}$ (specifically, taking $\replaceStar$ as the value of $j$ used to define $\parity_i$) such that
\begin{align}
\sum_{l=\replaceStar+\burst}^i \parity_l & = \label{eq:inductZero}\\
\sum_{l=\replaceStar+\burst}^{\replaceI+\delay} \parity_l & = \identity[\replaceStar+\burst-1 \ge \replaceI+1] \left(k_{\replaceI}- \policyShort_{\replaceI} \right) + \identity[\replaceStar \le \replaceI]\policyShort_{\replaceI} + \sum_{l = \replaceStar}^{\replaceI} (k_{l-1}- \policyShort_{l-1}') + \sum_{l = \replaceStar}^{\replaceI-1}\policyShort_{l}' \label{eq:inductFirst}\\
&= -\identity[\replaceStar = \replaceI+1]\policyShort_{\replaceStar-1} -\identity[\replaceStar > (\replaceI+1)]\policyShort_{\replaceStar-1}' -\identity[\replaceStar + \burst-1 =\replaceI] \left(k_{\replaceI} - \policyShort_{\replaceI}\right) + \sum_{l = \replaceStar-1}^{\replaceI} k_l \label{eq:inductSecond}\\
&= -\policyShort_{\replaceStar-1}' -\identity[\replaceStar + \burst-1 =\replaceI] \left(k_{\replaceI} - \policyShort_{\replaceI}\right) + \sum_{l = \replaceStar-1}^{\replaceI} k_l \label{eq:inductSecond2}\\
&= -\policyShort_{\replaceStar-1}' -\identity[\replaceStar + \burst-1 =i-\delay] \left(k_{i-\delay} - \policyShort_{i-\delay}\right) +\sum_{l = \replaceStar-1}^{i-\delay} k_l \label{eq:inductThird} \\
&\le -\policyShort_{\replaceStar-1} -\identity[\replaceStar + \burst-1 =i-\delay] \left(k_{i-\delay} - \policyShort_{i-\delay}\right) + \sum_{l = \replaceStar-1}^{i-\delay} k_l. \label{eq:inductFourth} 
\end{align}
\Equation~\eqref{eq:inductFirst} follows from the fact that $\replaceI=(i-\delay)$ and \Equation~\eqref{eq:paritySize}. \Equation~\eqref{eq:inductSecond} follows from rearranging terms. 
\Equation~\eqref{eq:inductSecond2} is immediate if $\replaceStar > \replaceI$ and otherwise follows from $\parity_{\replaceI+\delay} \le \policyShort_{\replaceI}$ (by $\replaceStar > \replaceI$ and \Equation~\eqref{eq:paritySize}) leading to $\policyShort_{\replaceI}' = \policyShort_{\replaceI}$ (by \Equation~\eqref{eq:adaptSpreadDef}). 
\Equation~\eqref{eq:inductThird} follows from substituting $\replaceI=(i-\delay)$. 
\Equation~\eqref{eq:inductFourth} follows from \Equation~\eqref{eq:adaptSpreadDef}. 

By Lemma~\ref{lem:worstCaseConstraint}, 
\begin{equation}
\label{eq:lowerBoundNumParityOther}
    \sum_{l=\replaceStar+\burst}^i \parityOther_l \ge -\policyShort_{\replaceStar-1}- \identity[\replaceStar +\burst-1 =i-\delay] \big(k_{i-\delay} - \policyShort_{i-\delay}\big)  + \sum_{l=\replaceStar-1}^{i-\delay} k_l.
\end{equation}
Combining \Equations~\eqref{eq:inductZero}, ~\eqref{eq:inductFourth}  and~\eqref{eq:lowerBoundNumParityOther} leads to
\begin{equation}
\label{eq:combineWithIH}
    \sum_{l=\replaceStar+\burst}^i \parityOther_l \ge \sum_{l=\replaceStar+\burst}^i \parity_l.
\end{equation}
Applying \Equation~\eqref{eq:IHConvert} (for $j = \replaceStar + \burst - 1$) to \Equation~\eqref{eq:combineWithIH} leads to 
\begin{equation}
\sum_{l=0}^{\replaceStar + \burst - 1} \parityOther_l + \sum_{l=\replaceStar + \burst - 1}^i \parityOther_l   = \sum_{l=0}^i \parityOther_l \ge \sum_{l=0}^{\replaceStar + \burst - 1} \parity_l + \sum_{l = \replaceStar + \burst - 1}^i \parity_l = \sum_{l=0}^i \parity_l,
\end{equation}
proving the inductive step for $l \in \{3\delay,\ldots,\len-\delay\}$. 
Recall that $\parity_{\len-\delay+1}=0,\ldots,\parity_\len = 0$, leading to
$$\sum_{l=0}^\len \parityOther_l \ge \sum_{l=0}^{\len-\delay} \parityOther_l \ge \sum_{l=0}^{\len-\delay} \parity_l = \sum_{l=0}^\len \parity_l.$$

The $\left(\delay, \burst,\len,\left \langle \policyShort_i \mid i \in [\len]\right\rangle\right)-$\spreadvgms sends $\sum_{l=0}^{\len} (k_l+\parity_l)$ symbols, which is no more than the number sent under the alternative construction (i.e., $\sum_{l=0}^{\len} (k_l+\parityOther_l)$).
\end{IEEEproof}

\section{Offline-optimal streaming codes}
\label{sec:offlineOptimal}
\begin{figure*}[t]
    \centering
\includegraphics[width=\linewidth]{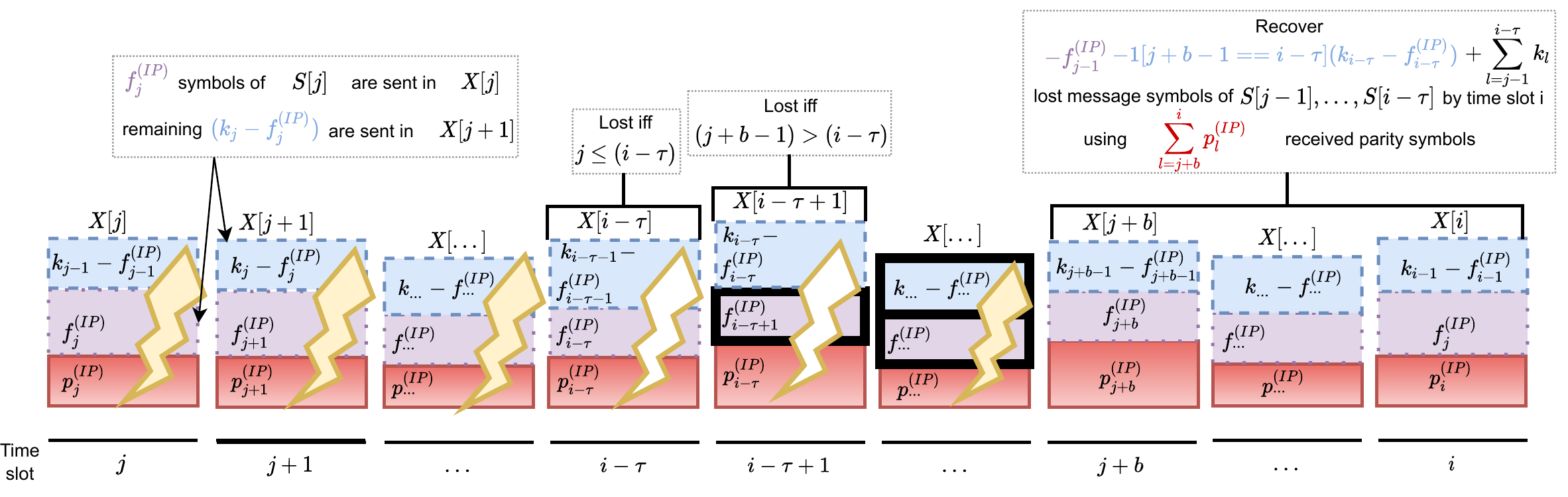}
    \caption{Modeling the transmission and constraints using the variables of the integer program. For any $i \in \{3\delay,\ldots,\len-\delay\}$ and burst (lightning bolts) of length $\burst$ starting in $j \in \{i-\delay-\burst+1,\ldots,i-\delay+1\}$, $\s[j-1],\ldots,\s[i-\delay]$, are recovered by time slot $j$ under the relaxation of receiving the lost symbols of $\s[i-\delay+1],\ldots,\s[j+\burst-1]$ (boxes with thick black outline).} 
    \label{fig:offline}
\end{figure*}

In this section, we design the first rate-optimal offline construction {for the setting of $\delay_L = 1$}. 
We build the construction with two steps for an arbitrary \messagePacketSizeSequence, $k_0,\ldots,k_\len$. 
First, we design an integer program (IP) to use constraints to model satisfying the \losslessDelay and Lemma~\ref{lem:worstCaseConstraint}. 
The IP determines the optimal policy for each time slot: $\left \langle \policyShort_i \mid i \in [\len] \right \rangle$, as is illustrated in Figure~\ref{fig:offline}. 
Second, we employ the $\left(\delay, \burst,\len,\left\langle \policyShort_i \mid i \in [\len]\right\rangle\right)-$\spreadvgms given the polices. 
The objective function of the integer program is to minimize the total number of parity symbols transmitted, which maximizes the rate.

{Next, we introduce Algorithm~\ref{alg:offlineOpt} to determine an optimal policy, $\policyShort_i$, for each time slot $i \in [\len]$ and then verify that the $\left(\delay, \burst,\len,\left\langle \policyShort_i \mid i \in [\len]\right\rangle\right)-$\spreadvgms is rate optimal.}
\begin{alg}
\caption{Computes $\left \langle \policyShort_i \mid i \in [\len] \right \rangle$ for which the $\left(\delay, \burst,\len,\left\langle \policyShort_i \mid i \in [\len]\right\rangle\right)-$\spreadvgms matches the \offlineOptimal.}
\label{alg:offlineOpt}
\noindent \textbf{Input:} $(\delay, \burst, \len, k_0,\ldots,k_\len)$

Minimize $\sum_{i=0}^{\len} \parityOffline_i$ subject to
\begin{itemize}
\item $\forall i \in [\len]$, $\policyOffline_i \ge 0,  \policyOffline_i \le k_i, \parityOffline_i \ge 0$.
\item $\forall i \in \{3\delay, \ldots, \len-\delay\},j \in  \{i-\delay-\burst+1,\ldots,i-\delay+1\}$, 
\end{itemize}
\begin{equation*}
-\policyOffline_{j-1}- \identity[j+\burst-1 =i-\delay]\left(k_{i-\delay}-\policyOffline_{i-\delay}\right) + \sum_{l=j-1}^{i-\delay} k_l \le \sum_{l=j+\burst}^{i} \parityOffline_{l}.
\end{equation*}
\noindent \textbf{Output:} $\left \langle \policyShort_i \mid i \in [\len] \right \rangle$ 
\end{alg}
\begin{theorem}
\label{lem:remaining}
{For any $(\delay, \burst,\len)$ and \messagePacketSizeSequence $k_0,\ldots,k_\len$, suppose Algorithm~\ref{alg:offlineOpt} outputs $\left \langle \policyShort_i \mid i \in [\len] \right \rangle$. Then the $\left(\delay, \burst,\len,\left\langle \policyShort_i \mid i \in [\len]\right\rangle\right)-$\spreadvgms is rate optimal.} 
\end{theorem}
\begin{IEEEproof}
{Due to Lemma~\ref{lem:changeSameRate}, the rate of the $\left(\delay, \burst,\len,\left\langle \policyShort_i \mid i \in [\len]\right\rangle\right)-$\spreadvgms is the same as a construction that for $i \in [\len]$ employs the policy $\policyShort_i$ and sends $\parityOffline_i$ parity symbols in $\x[i]$. 
We will show that no coding scheme can send fewer than $\sum_{i=0}^{\len} \parityOffline_i$ parity symbols.} 

{An arbitrary rate-optimal construction must satisfy the first constraint because for all $i \in [\len]$ between $0$ and $k_i$ symbols of $\s[i]$ are sent in $\x[i]$ along with a non-negative number of parity symbols. 
The construction must satisfy the second constraint due to Lemma~\ref{lem:worstCaseConstraint}. 
Using each policy of this rate-optimal construction along with the number of parity symbols it sends is a valid solution to the integer program. 
Correctness follows from minimization.} 
\end{IEEEproof}

Although Algorithm~\ref{alg:offlineOpt} applies to the entire \messagePacketSizeSequence, it is trivial to modify the algorithm to apply to the remainder of a transmission after \channelPackets $\x[0],\ldots,\x[l]$ have been sent for some $l \in [\len]$. 
This involves adding constraints for all $j \in [l]$ (a) $\policyOffline_j = \policyShort_j$ and (b) $\parityOffline_j = \parity_j$. 
We call the modified algorithm ``\algModified.'' 
\begin{corollary}
\label{cor:offlinePartialOptimal}
{For any $(\delay, \burst,\len)$, \messagePacketSizeSequence $k_0,\ldots,k_\len$, and $l \in [\len-\delay]$, suppose that for all $j \in [l]$, policy $\policyShort_j$ was used and $p_{j}$ parity symbols were sent in $\x[j]$, and \algModified outputs $\left \langle \policyShort_i \mid i \in [\len] \right \rangle$. Then the $\left(\delay, \burst,\len,\left\langle \policyShort_i \mid i \in [\len]\right\rangle\right)-$\spreadvgms attains the best possible rate given the prior transmission of $\x[0],\ldots,\x[l]$.} 
\end{corollary}

\section{Learning-based online streaming codes}
\label{sec:onlineOptimal}
\begin{figure*}[t]
    \centering
\includegraphics[width=.25\linewidth]{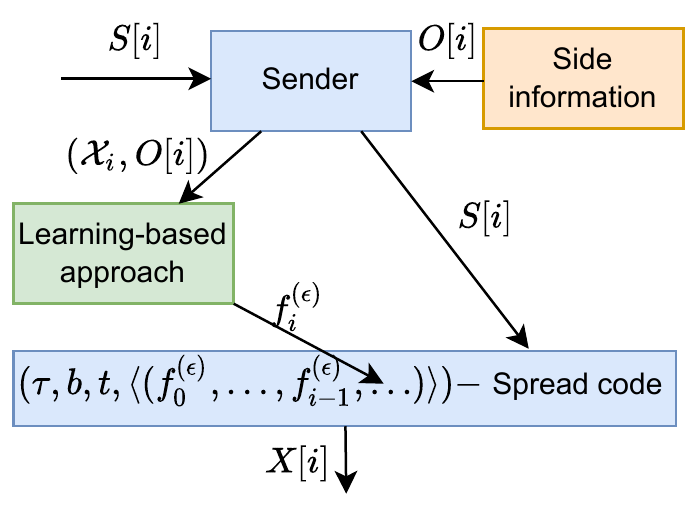}
    \caption{Illustration of the $(\delay, \burst,\len)-$\learningCode. A learning-based approach is used to determine a policy, $\policyPredictShort_i$, during the $i$th time slot, which is then used by the  $\left(\delay, \burst,\len,\left\langle \policyPredictShort_i \mid i \in [\len]\right\rangle\right)-$\spreadvgms.} 
    \label{fig:online}
\end{figure*}

We now present an online code construction, dubbed the ``$(\delay, \burst,\len)-$\learningCode,'' whose expected rate is within $\epsilon$ of the \onlineOptimal. 
The construction uses a learning-based approach to specify the policy of spreading the symbols of $\s[i]$, denoted $\policyPredictShort_i$, for each $i \in [\len]$, and then applies the $\left(\delay, \burst,\len,\left\langle \policyPredictShort_i \mid i \in [\len]\right\rangle\right)-$\spreadvgms, as is shown at a high level in Figure~\ref{fig:online}. 

\begin{figure*}[t]
    \centering
\includegraphics[width=.5\linewidth]{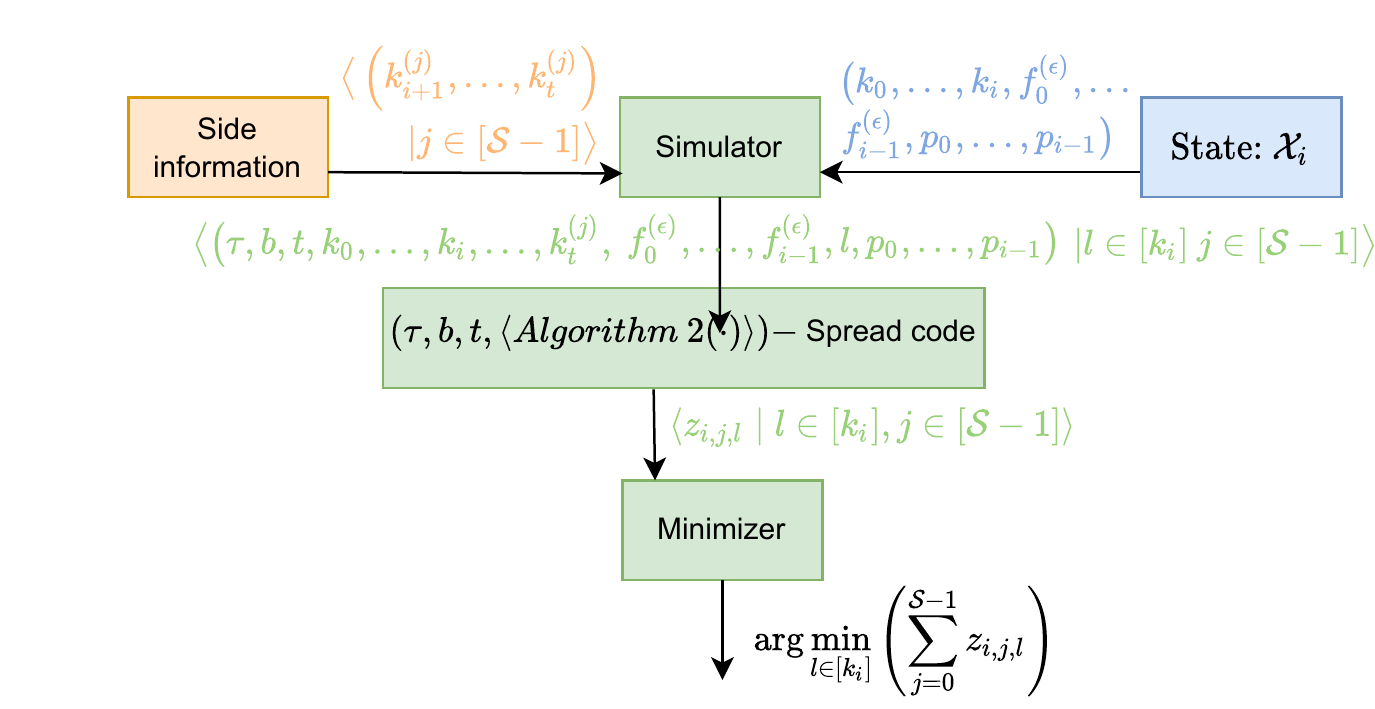}
    \caption{Illustration of the learning-based approach (green) to determine how to spread message symbols.}
    \label{fig:predictor}
\end{figure*}

To determine how to spread message symbols, 
we use the \oracleInfo of $\sample$ samples the {distribution of} the sizes of the future \messagePackets, which was defined in \Equation~\eqref{eq:defOracle} as $\big \langle (k_{i+1}^{(j)},\ldots,k_{\len}^{(j)}) \sim D_{k_0,\ldots,k_i} \mid j \in \{0,\ldots,\sample-1\} \big \rangle$. 
We use a similar technique to empirical risk minimization over the $\sample$ samples to set $\policyPredictShort_i$ to the value leading to lowest expected number of symbols being sent by a rate-optimal offline code. 
Specifically, for any $i = 0,\ldots,\len$, $j \in [\sample-1]$ and $l \in [k_i]$, let 
\begin{equation*}
\regretVariable_{i,j,l} = \sum_{r=i}^\len \parityOffline_r,    
\end{equation*}
where $\parityOffline_i,\ldots,\parityOffline_\len$ are variables used by the IP of \algModified given $\x[0],\ldots,\x[i-1],$ and $\policyOffline_i = l$. 
Then
\begin{equation}
\label{eq:defPredict}
 \policyPredictShort_i = \arg \min_{l \in [k_i]} \frac{1}{\sample-1} \sum_{j \in [\sample-1]} \regretVariable_{i,j,l}. 
\end{equation}
We demonstrate how $\policyPredictShort_i$ is defined in Figure~\ref{fig:predictor}. 

{The key observation to interpret the choice of $\policyPredictShort_i$ is that the number of parity symbols sent corresponding to \messagePacket $\s[i]$, namely $\parity_{i+\delay}$, is monotonically non-decreasing as $\policyPredictShort_i$ increases.} 
{Thus, smaller values of $\policyPredictShort_i$ lead to smaller values of $\parity_{i+\delay}$, which exploits the parity symbols already sent before time slot $(i+\delay)$. 
This strategy is effective when the next several \messagePackets are likely small. 
Therefore, a small $\p[i+\delay]$ suffices to ensure that the next several \messagePackets are recovered when some of $\x[i+2],\ldots,\x[i+\delay-1]$ are lost. 
In contrast, larger values of $\policyPredictShort_i$ promote larger values of $\parity_{i+\delay}$, which is suitable when the next several \messagePackets are likely to be large. Hence, a large $\p[i+\delay]$ will not go to waste even if a burst starts after receiving $\s[i]$.}

{To show that the $(\delay, \burst,\len)-$\learningCode is approximately rate optimal, we analyze the number of extra symbols it sends compared to an optimal scheme as follows. }
\begin{definition}[Regret]
\label{def:regret}
The \textit{regret, $\regret_{k_i,\ldots,k_\len}\left(\policyPredictShort_i\right)$}, for the \messagePacketSizeSequence $k_0,\ldots,k_\len$ is the number of extra symbols sent under \algModified when $\policyPredictShort_i$ is used compared to the best offline policy, $\policyShort_i'$.
\end{definition}

The number of extra symbols sent compared to an optimal offline scheme is {$\sum_{i=0}^\len \regret_{k_i,\ldots,k_\len}\left(\policyPredictShort_i\right)$}, as is shown next for completeness.
\begin{lemma}
For \messagePacketSizeSequence $k_0,\ldots,k_\len$, the $(\delay, \burst,\len)-$\learningCode transmits $\sum_{i=0}^\len \regret_{k_i,\ldots,k_\len}\left(\policyShort_i\right)$ more symbols than a scheme meeting the \offlineOptimal.
\label{lem:regret}
\end{lemma}
\begin{IEEEproof}
We can sequentially improve the  $(\delay, \burst,\len)-$\learningCode for $i = \len-2\delay, \ldots,2\delay$ by switching $\policyPredictShort_j$ for $j \in \{i,\ldots,\len\}$ to the one computed by \algModified given $\policyPredictShort_0,\ldots,\policyPredictShort_{i-1}, \parity_0,\ldots,\parity_{i-1}$. 
For each value of $i$, the improvement in total number of symbols sent is $\regret_{k_i,\ldots,k_\len}\left(\policyShort_i\right)$ by Definition~\ref{def:regret}. 
After reaching $i = 2\delay$, the output is simply \algModified, as $k_0=0,\ldots,k_{2\delay-1} = 0$. 
\end{IEEEproof}

Next, we bound the expected regret of the $(\delay, \burst,\len)-$\learningCode from spreading message symbols for any time slot.
\begin{lemma}
\label{lem:regretPerTimeslot}
For any $(\delay, \burst, \len)$, $\sample \ge \sqrt{\ln(\frac{8\maxSize^2}{\epsilon})}\frac{2\sqrt{2}\maxSize^3}{\epsilon}$ samples from the \oracleInfo, where $\maxSize$ is the maximum size of a \messagePacket, $i \in [\len]$, $k_0,\ldots,k_{i-1},$ and $\policyShort_i' \in [k_i]$, 
\begin{equation}
\label{eq:predictionAccuracy}
    \E_{k_{i+1},\ldots,k_{\len}}\left[\regret_{k_i,\ldots,k_\len}\left(\policyPredictShort_i\right) -\regret_{k_i,\ldots,k_\len}\left( \policyShort_i'\right)\right] \le \epsilon.
\end{equation}
\end{lemma}
\begin{IEEEproof}
If $k_i=0$, $\policyShort_i = \policyPredictShort_i$, concluding the proof. 
Otherwise, the choice of $\policyPredictShort_i$ leads to sending $k_i$ extra parity symbols in $\x[i+\delay]$ compared to an optimal scheme, so 
\begin{equation}
\label{eq:boundRegret}
\regret_{k_i,\ldots,k_\len}\left(\policyPredictShort_i\right) \le \maxSize
\end{equation} and 
\begin{equation*}
    \E_{k_{i+1},\ldots,k_{\len}}\left[\regret_{k_i,\ldots,k_\len}\left(\policyPredictShort_i\right)\right] \le \maxSize, \: \var_{k_{i+1},\ldots,k_{\len}}\left(\regret_{k_i,\ldots,k_\len}\left(\policyPredictShort_i\right)\right) \le \maxSize^2.
\end{equation*}

At a high level, we apply the Hoeffding bound~\cite{hoeffding1994probability} to show that the expected regret for each possible policy is well approximated using the empirical mean over the $\sample$ samples from the \oracleInfo. 
The unlikely event that the expected value deviates greatly from the mean will have negligible impact due to \Equation~\eqref{eq:boundRegret}.

Next, we use $\oracle[i]$ to determine the values of $\sample$ random variables, $\left(\regretVariable_{i,j,0},\ldots,\regretVariable_{i,j,\sample-1}\right)$, equaling $\regret_{k_i,\ldots,k_\len}(j)$ in distribution. 
The empirical mean, $\frac{1}{\sample}\sum_{l=0}^{\sample-1}\regretVariable_{i,j,l}$, is used to estimate $\E_{k_i,\ldots,k_\len}[\regret_{k_i,\ldots,k_\len}(j)]$. 
By the Hoeffding bound~\cite{hoeffding1994probability}, 
\begin{equation*}
\left|\frac{1}{\sample}\left(\sum_{l=0}^{\sample-1} \regretVariable_{i,j,l}\right) - \E_{k_0,\ldots,k_\len}[\regret_{k_i,\ldots,k_\len}(j)]\right| < \epsilon_{\dagger}    
\end{equation*}
 with probability at least 
 \begin{equation*}
     \left(1-2 e^{-\frac{2\sample^2(\epsilon_{\dagger})^2}{\maxSize^2}}\right) \ge (1-\delta)
\end{equation*} as long as 
\begin{align*}
\frac{\delta}{2} &\ge e^{-\frac{2\sample^2(\epsilon_{\dagger})^2}{\maxSize^2}}\\ 
\frac{2\sample^2(\epsilon_{\dagger})^2}{\maxSize^2} &\ge \ln\left(\frac{2}{\delta}\right)\\
\sample^2 &\ge \ln\left(\frac{2}{\delta}\right) \frac{\maxSize^2}{2(\epsilon_{\dagger})^2} \\
\sample & \ge \sqrt{\ln\left(\frac{2}{\delta}\right)}\frac{\maxSize}{\sqrt{2}\epsilon_{\dagger}}.    
\end{align*}

Using $\epsilon_{\dagger} = \delta = \frac{\epsilon}{4\maxSize^2}$ and applying the union bound over the at most $\maxSize$ values of $k_i$ shows with probability $(1-\maxSize \delta)$  for all $j \in [k_i]$, 
\begin{equation}
\label{eq:union}
\left|\frac{1}{\sample}\left(\sum_{l=0}^{\sample-1} \regretVariable_{i,j,l}\right) - \E_{k_0,\ldots,k_\len}\left[\regret_{k_i,\ldots,k_\len}\right]\left(\policyPredictShort_{i,j}\right)]\right| < \epsilon_{\dagger}.    
\end{equation}
We set $\policyPredictShort_{i}$ according to \Equation~\eqref{eq:defPredict} as
\begin{equation*}
    \policyPredictShort_{i} = \arg \min_{j} \frac{1}{\sample} \sum_{l=0}^{\sample-1} \regretVariable_{i,j,l}.
\end{equation*}

With probability $(1-\maxSize \delta)$ \Equation~\eqref{eq:union} holds, leading to
\begin{equation}
\label{eq:firstRegret}
\begin{aligned}
\E_{k_i,\ldots,k_\len}\left[\regret_{k_i,\ldots,k_\len}\left(\policyPredictShort_{i}\right) -\regret_{k_i,\ldots,k_\len}\left( \policyShort_i'\right)\right] \le& \\
\E_{k_i,\ldots,k_\len}\left[\regret_{k_i,\ldots,k_\len}\left(\policyPredictShort_{i}\right) -\regret_{k_i,\ldots,k_\len}\left( \policyShort_i'\right)\right] + \left(\frac{1}{\sample}\left(\sum_{l=0}^{\sample-1} \regretVariable_{i,\policyShort_i',l}\right) - \frac{1}{\sample}\left(\sum_{l=0}^{\sample-1} \regretVariable_{i,\policyPredictShort_{i},l}\right)\right) \le& \\
 \left (\E_{k_i,\ldots,k_\len}\left[\left(\regret_{k_i,\ldots,k_\len}\left(\policyPredictShort_{i}\right)\right]- \frac{1}{\sample}\left(\sum_{l=0}^{\sample-1} \regretVariable_{i,\policyPredictShort_{i},l}\right) \right) \right) + \left(\left(\frac{1}{\sample}\left(\sum_{l=0}^{\sample-1} \regretVariable_{i,\policyShort_i',l}\right)-\E_{k_i,\ldots,k_\len}\left[\regret_{k_i,\ldots,k_\len}\left( \policyShort_i'\right)\right] \right)\right ) \le\\
 \left |\E_{k_i,\ldots,k_\len}\left[\regret_{k_i,\ldots,k_\len}\left(\policyPredictShort_{i}\right)\right]- \frac{1}{\sample}\left(\sum_{l=0}^{\sample-1} \regretVariable_{i,\policyPredictShort_{i},l}\right) \right| + \left|\left(\frac{1}{\sample}\left(\sum_{l=0}^{\sample-1} \regretVariable_{i,\policyShort_i',l}\right)-\E_{k_i,\ldots,k_\len}\left[\regret_{k_i,\ldots,k_\len}\left( \policyShort_i'\right)\right] \right)\right | \le&2\epsilon_{\dagger},
\end{aligned}
\end{equation} 
which used the fact that \Equation~\eqref{eq:defPredict} led to
$$\frac{1}{\sample}\left(\sum_{l=0}^{\sample-1} \regretVariable_{i,\policyShort_i',l}\right) \ge \frac{1}{\sample}\left(\sum_{l=0}^{\sample-1} \regretVariable_{i,\policyPredictShort_{i},l}\right).$$

Otherwise, with probability $\maxSize \delta$
\begin{equation}
    \label{eq:secondRegret}
\E_{k_i,\ldots,k_\len}\left[\regret_{k_i,\ldots,k_\len}(\policyPredictShort_{i}) -\regret_{k_i,\ldots,k_\len}\left( \policyShort_i'\right)\right]
    \le \maxSize.
\end{equation}
Combining \Equations~\eqref{eq:firstRegret} and~\eqref{eq:secondRegret} leads to 
\begin{align*}
    \E_{k_i,\ldots,k_{\len}}\left[\regret_{k_i,\ldots,k_\len}(\policyPredictShort_{i}) -\regret_{k_i,\ldots,k_\len}\left( \policyShort_i'\right)\right] 
    \le \delta \maxSize^2 + 2\epsilon_{\dagger} \le \epsilon.
\end{align*}
\end{IEEEproof}

Finally, we show that the expected \onlineOptimal, denoted as ``$\rateOptimal$,'' is within $\epsilon$ of the expected rate of the $(\delay, \burst,\len)-$\learningCode, denoted as
\begin{equation}
    \label{eq:expRate}
    \rate = \E_{k_0,\ldots,k_\len} \left[ \frac{\sum_{i=0}^\len k_i}{\sum_{i=0}^\len k_i + \parity_i}\right].
\end{equation}

\begin{theorem}
For any $(\delay, \burst, \len)$ and for $\sample \ge \sqrt{\ln(\frac{8\maxSize^2}{\epsilon})}\frac{2\sqrt{2}\maxSize^3}{\epsilon}$ samples from the \oracleInfo, where $\maxSize$ is the maximum size of a \messagePacket, $(\rateOptimal-\rate) < \epsilon$. 
\label{thm:onlineOpt}
\end{theorem}
\begin{IEEEproof}
{We consider an online scheme with the optimal expected rate, namely the $\left(\delay, \burst,\len,\left \langle \policyShort_i' \mid i \in [\len]\right\rangle\right)-$\spreadvgms, which must exist by Lemma~\ref{lem:changeSameRate}.} 
{Let $n_i$ be the number of symbols sent in $\x[i]$ under the optimal scheme and } 
\begin{equation}
\label{eq:totalRegret}
\begin{aligned}
 n_\epsilon&=\sum_{i=0}^\len \regret_{k_i,\ldots,k_\len}\left(\policyPredictShort_i\right) -\regret_{k_i,\ldots,k_\len}\left( \policyShort_i'\right)  
\end{aligned}
\end{equation}
be the number of additional symbols sent under the $\left(\delay, \burst,\len,\left \langle \policyShort_i' \mid i \in [\len]\right\rangle\right)-$\spreadvgms. 

By definition
\begin{align}
\rateOptimal-\rate &\le \E_{k_0,\ldots,k_\len} \left[\left |\frac{\sum_{i=0}^\len k_i}{\sum_{i=0}^\len n_i}- \frac{\sum_{i=0}^\len k_i}{n_\epsilon + \sum_{i=0}^\len n_i} \right|\right] \label{eq:onlineOverall} \\
&\le \E_{k_0,\ldots,k_\len} \left[ \left |\frac{|n_\epsilon|\sum_{i=0}^\len k_i}{\left(\sum_{i=0}^\len n_i\right)\left(n_\epsilon + \sum_{i=0}^\len n_i\right)} \right| \right] \nonumber\\
&\le \E_{k_0,\ldots,k_\len} \left[
\frac{|n_\epsilon|}{\sum_{i=0}^\len n_i} \right]\label{eq:proofEquationOnlineFirst} \\
&\le \E_{k_0,\ldots,k_\len} \left[
\frac{|n_\epsilon|}{\sum_{i=0}^\len \identity[k_i > 0] }\right], \label{eq:proofEquationOnlineSecond} 
\end{align}
where \Equation~\eqref{eq:proofEquationOnlineFirst} follows from sending $\sum_{i=0}^\len n_i \ge \sum_{i=0}^\len k_i$ symbols to satisfy the \losslessDelay constraint, and \Equation~\eqref{eq:proofEquationOnlineSecond} follows from $\sum_{i=0}^\len k_i \ge \sum_{i=0}^\len \identity[k_i>0]$.

If $k_i = 0$, then $\regret_{k_i,\ldots,k_\len}\left(\policyPredictShort_i\right)  = \regret_{k_i,\ldots,k_\len}\left( \policyShort_i'\right) = 0$. 
Thus, we can simplify \Equation~\eqref{eq:totalRegret} as
\begin{equation}
\label{eq:totalRegretSimplified}
n_\epsilon=\sum_{i=0}^\len \identity[k_i > 0]\left(\regret_{k_i,\ldots,k_\len}\left(\policyPredictShort_i\right) -\regret_{k_i,\ldots,k_\len}\left( \policyShort_i'\right)\right).    
\end{equation}

Applying \Equation~\eqref{eq:totalRegretSimplified} to \Equations~\eqref{eq:onlineOverall} and~\eqref{eq:proofEquationOnlineSecond} leads to
\begin{align*}
    \rateOptimal-\rate &\le \E_{k_0,\ldots,k_\len} \left[
\frac{\sum_{i=0}^\len \identity[k_i > 0]\left( \regret_{k_i,\ldots,k_\len}\left(\policyPredictShort_i\right) -\regret_{k_i,\ldots,k_\len}\left( \policyShort_i'\right) \right)}{\sum_{i=0}^\len \identity[k_i > 0] }\right]  \\
&\le \max_{i \in [\len]} \E_{k_0,\ldots,k_\len} \left[\regret_{k_i,\ldots,k_\len}\left(\policyPredictShort_i\right) -\regret_{k_i,\ldots,k_\len}\left( \policyShort_i'\right) \right]\\
&=  \max_{i \in [\len]} \E_{k_0,\ldots,k_{i-1}} \left[\E_{k_i,\ldots,k_{\len}} \left[\regret_{k_i,\ldots,k_\len}\left(\policyPredictShort_i\right) -\regret_{k_i,\ldots,k_\len}\left( \policyShort_i'\right) \right]\right]
\end{align*}

As such, it suffices to show for all $i \in [\len]$ that 
\begin{equation}
\label{eq:perPacketRegret}
    \E_{k_0,\ldots,k_{i-1}}\left[\E_{k_i,\ldots,k_{\len}}\left[\regret_{k_i,\ldots,k_\len}\left(\policyPredictShort_i\right) -\regret_{k_i,\ldots,k_\len}\left( \policyShort_i'\right)\right]\right] \le \epsilon.
\end{equation}

Because $\sample \ge \sqrt{\ln(\frac{8\maxSize^2}{\epsilon})}\frac{2\sqrt{2}\maxSize^3}{\epsilon}$, Lemma~\ref{lem:regretPerTimeslot} guarantees for any $k_0,\ldots,k_{i-1}$, 
$$\E_{k_i,\ldots,k_{\len}}\left[\regret_{k_i,\ldots,k_\len}\left(\policyPredictShort_i\right) -\regret_{k_i,\ldots,k_\len}\left( \policyShort_i'\right)\right] \le \epsilon,$$
concluding the proof.
\end{IEEEproof}

We have designed an online code that uses a black box algorithm to determine how to spread message symbols and bounded how close the rate is to optimal based on the regret due to the choices of how to spread. To show that the code is approximately rate optimal, we presented an explicit learning-based approach of leveraging samples to the distribution of the sizes of future \messagePackets to spread message symbols (i.e., \Equation~\eqref{eq:defPredict}) and showed in Lemma~\ref{lem:regretPerTimeslot} that it has a sufficiently small expected regret. More generally, any criteria with a sufficiently small expected regret could be used, leading to the following result.

\begin{corollary}
Theorem~\ref{thm:onlineOpt} holds when any criteria for spreading message symbols is substituted for \Equation~\eqref{eq:defPredict} if the criteria satisfies \Equation~\eqref{eq:predictionAccuracy} for all $i \in [\len]$, $k_0,\ldots,k_\len,$ and $\policyShort_i' \in [k_i]$.
\end{corollary}

\section{Conclusion}
\label{sec:conclusion}

Inspired by the growing field of learning-augmented algorithms, this work introduces a new methodology for constructing online streaming codes that combines machine learning with algebraic coding theory tools. The approach is to (a) isolate the component that can benefit from machine learning, (b) solve the offline version of the problem by integrating optimization with algebraic coding theory techniques, and (c) convert the offline scheme into an online one using a learning-based approach. This strategy is applicable beyond the setting considered in this paper, including numerous other settings for real-time streaming communication.

\section*{Acknowledgment}
This work was funded in part by an NSF grant (CCF-1910813).


\bibliographystyle{IEEEtran}
\bibliography{the_bib}

\end{document}